\documentclass[10pt,twocolumn,letterpaper]{article}

\usepackage{cvpr}              

\definecolor{cvprblue}{rgb}{0.21,0.49,0.74}
\usepackage[pagebackref,breaklinks,colorlinks,allcolors=cvprblue]{hyperref}
\usepackage{amsmath,amssymb,amsfonts}
\usepackage{algorithmic}
\usepackage{graphicx}
\usepackage{appendix}
\usepackage{textcomp}
\usepackage{xcolor}
\usepackage{makecell}
\usepackage{float}
\usepackage{placeins}
\usepackage{multirow}
\usepackage{caption}
\usepackage{hyperref} 
\usepackage{tikz}
\usepackage{subcaption}
\usepackage{booktabs}
\usepackage{subcaption}
\usepackage{amssymb}
\usepackage{footmisc}
\def\paperID{12442} 
\def\confName{CVPR}
\def\confYear{2025}

\title{SMTPD: A New Benchmark for Temporal Prediction of Social
Media Popularity}

\author{
\textsuperscript{1}Yijie Xu\textsuperscript{*},
\textsuperscript{1}Bolun Zheng\textsuperscript{*}\textsuperscript{\dag},
\textsuperscript{1}Wei Zhu\textsuperscript{*},
\textsuperscript{1}Hangjia Pan,
\textsuperscript{1}Yuchen Yao,\\
\textsuperscript{2}Ning Xu,
\textsuperscript{2}Anan Liu,
\textsuperscript{3}Quan Zhang,
\textsuperscript{1}Chenggang Yan\\ 
\textsuperscript{1}Hangzhou Dianzi University, 
\textsuperscript{2}Tianjin University,
\textsuperscript{3}Peking University
}

\begin{document}

\maketitle
\footnotetext[1]{These authors contributed equally to this work.}
\footnotetext[2]{Corresponding author} 
\footnotetext{This work is supported by the the National Key Research and Development Program of China under Grant 2020YFB1406600.}

\begin{abstract}
Social media popularity prediction task aims to predict the popularity of posts on social media platforms, which has a positive
driving effect on application scenarios such as content optimization, digital marketing and online advertising. Though many studies
have made significant progress, few of them pay much attention
to the integration between popularity prediction with temporal
alignment. In this paper, with exploring YouTube’s multilingual
and multi-modal content, we construct a new social media temporal
popularity prediction benchmark, namely SMTPD, and suggest a
baseline framework for temporal popularity prediction. Through
data analysis and experiments, we verify that temporal alignment
and early popularity play crucial roles in social media popularity
prediction for not only deepening the understanding of temporal
dynamics of popularity in social media but also offering a suggestion about developing more effective prediction models in this
field. Code is available at \url{https://github.com/zhuwei321/SMTPD}
\end{abstract}
\vspace{-1cm}   
\section{Introduction}

With the advancement of Internet communication technology in recent years, social media has gradually emerged and has influenced various aspects of human life. 
Any content posted on social media stands the chance of becoming hot spot, and widely disseminated social media content can generate significant social and economic benefits. 
The prediction of social media popularity holds immense potential applications in content optimization \cite{agarwal2008online,lifshits2010ediscope}, online advertising \cite{ghose2009empirical,gharibshah2021user}, digital marketing \cite{hajarian2021taxonomy,tatar2014survey}, search recommendations \cite{gonccalves2010popularity,bao2007optimizing}, intelligent fashion \cite{cheng2021fashion}, and beyond.
\par

\begin{figure}[t]
    \centering
    \subfloat[A social media post, also serving as a sample in SMTPD.]{
        \includegraphics[width=1\linewidth]{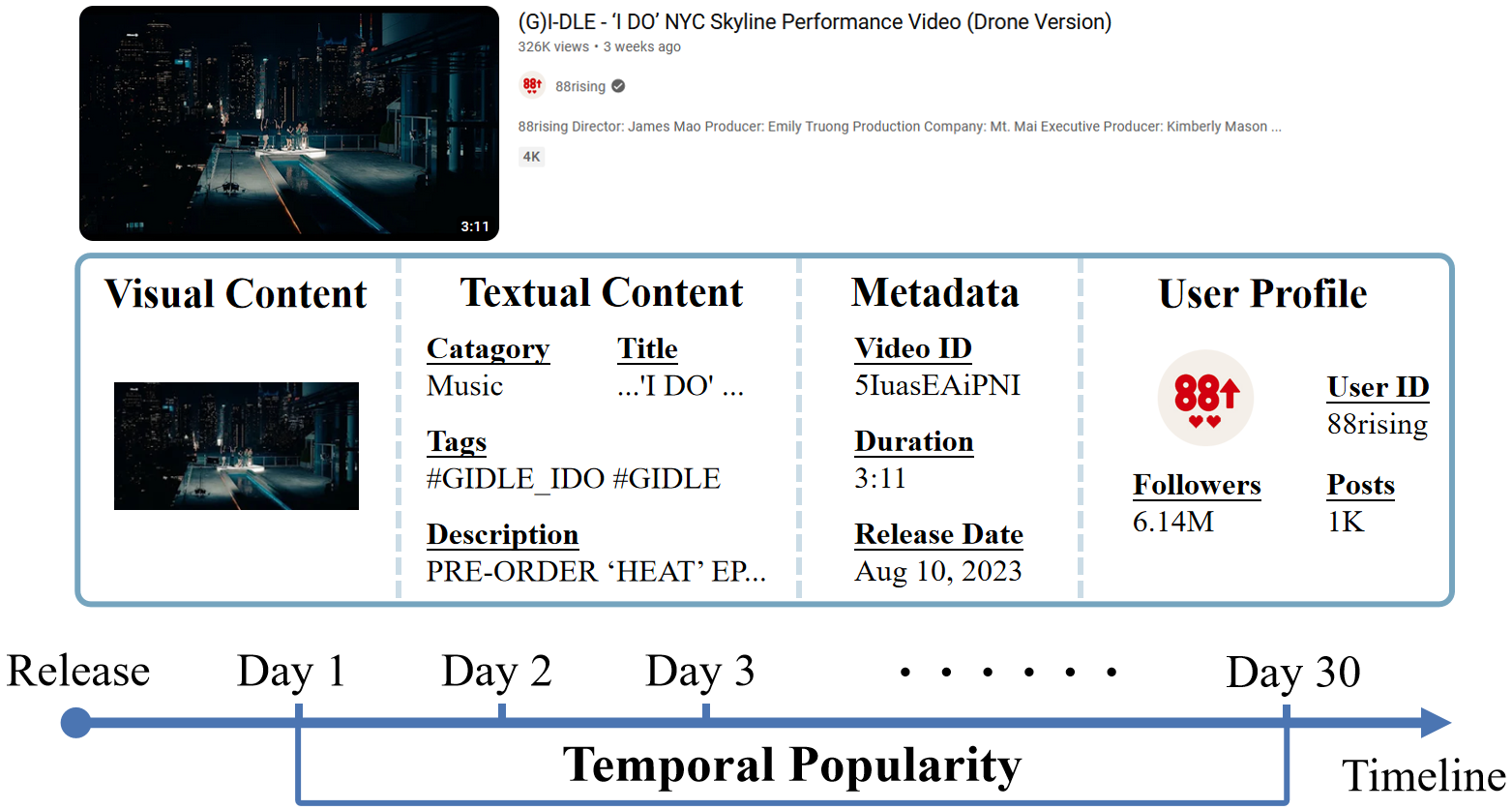}
        \label{fig:sample}
    }
    \\
    \subfloat[Box plots of popularity scores over time.]{
        \includegraphics[width=1\linewidth]{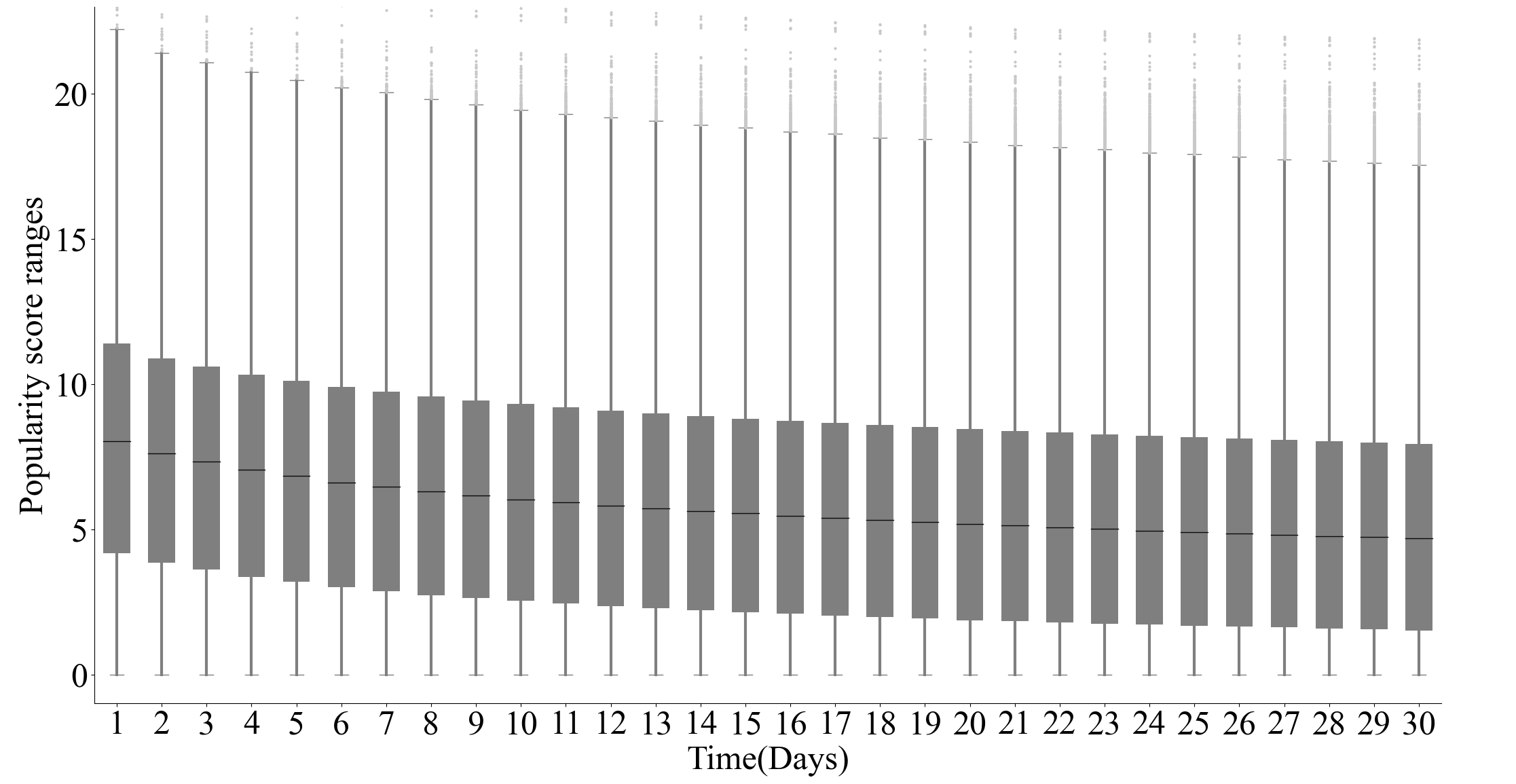}
        \label{fig:box}
    }
    \caption{Content sections and popularity trend of SMTPD. In \ref{fig:sample}, a sample involves four sections of content, with temporal popularity. \ref{fig:box} depicts box plots of daily popularity scores, illustrating variations in popularity distribution at different time points. The distribution consistently demonstrates a decay pattern over time.}
    \label{fig:distribution}
\end{figure}

In the early stages of popularity research, statistical and topological methods \cite{szabo2010predicting, yang2011patterns, richier2014bio} were extensively employed, with the research focus primarily on time-aware popularity prediction. 
These prediction methods rely on information about the earlier popularity of posted content.
Along with the many large-scale social media popularity datasets are proposed \cite{bao2013popularity, sanjo2017recipe, wu2017sequential, wu2019smp}, the machine learning based methods are widely employed to achieve reasonable predictions of social media popularity and have shown remarkable performance.
These methods emphasize meticulous design in feature extraction and fusion, followed by popularity prediction using machine learning models such as deep neural networks (DNN) and gradient boosting decision trees (GBDT).
\par

Table \ref{table:compare} lists the public or mentioned social media popularity datasets in the most recent years as far as we know. 
For above machine learning based methods, the large-scale social media popularity datasets \cite{bao2013popularity, mazloom2016multimodal, sanjo2017recipe, wu2017sequential, wu2019smp, cho2024amps} provide the fundamental supports.

However, existing datasets and related studies have notable limitations, such as insufficient multi-modal data, limited language diversity, and, crucially, the absence of a consistent timeline for prediction. As shown in Figure\ref{fig:box}, popularity distributions shift over time, and predictions made at irregular intervals can reduce accuracy and create challenges for practical applications.

\begin{table}[t] 
\centering

\renewcommand\arraystretch{1.5}
\setlength{\tabcolsep}{4pt}
\resizebox{\columnwidth}{!}{%
\begin{tabular}{@{\extracolsep{\fill}} cccccc} 
\toprule
\textbf{Dataset}  & \textbf{Source} & \textbf{Category} & \textbf{Samples} & \textbf{Language} & \textbf{Prediction Type}\\
\midrule
Mazloom \cite{mazloom2016multimodal} & Instagram & fast food brand & 75K & English & single\\
Sanjo \cite{sanjo2017recipe} & Cookpad & recipe & 150K & Japanese & single\\
TPIC17 \cite{wu2017sequential} & Flickr & - & 680K & English & single\\
SMPD \cite{wu2019smp} & Flickr & 11 categories & 486K & English & single\\
AMPS \cite{cho2024amps} & YouTube & shorts & 13K & Korean & single \\
\midrule
SMTPD (ours) & YouTube & 15 categories & 282K & over 90 languages & sequential\\
\midrule
\end{tabular}%
}
\caption{Comparing SMTPD with existed multi-modal social media popularity datasets.}
\label{table:compare}
\end{table}

Furthermore, effective popularity prediction requires integrating both the multi-modal content of social media and information on the communication process to capture the dynamic trends in popularity. This essential aspect is missing in most current prediction efforts.
\par
Targeting to fix up the above shortcomings of existed datasets,
in this paper, we propose a new benchmark called Social Media Temporal Popularity Dataset (SMTPD) by observing multi-modal content from mainstream for cutting-edge research in the field of social multimedia along with multi-modal feature temporal prediction. 
As shown in Figure \ref{fig:sample}, multi-modal contents of social media posts in multiple languages generated the daily popularity information during the communication process. We define such a post as a sample of SMTPD, and
propose a multi-modal framework as a baseline to achieve the temporal prediction of popularity scores.The proposed framework consists of two parts, feature extraction and regression. 
In the feature extraction part, multiple pre-training models and pre-processing methods are introduced to translate the multi-modal content into deep features. 
In the regression part, we encode the state and time sequence of the extracted features, and adopt the LSTM-based structure to regress the 30-day popularities.
Through analysis and experiments, we discover the importance of early popularity to the task of popularity prediction, and demonstrate the effectiveness of our method in temporal prediction. 
Generally, the contribution of this work can be summarized as:
\begin{itemize}
    \item Against the missing of temporal information in social media popularity researches, we observe over 282K multilingual samples from mainstream social media since they released on the network lasting for 30 days. We refer to these samples as SMTPD, a new benchmark for temporal popularity prediction.
    \item Basing on existed methods, we innovate in both the selection of feature extractors and the construction of the temporal regression component, and suggest a baseline model which enables temporal popularity prediction to be conducted across multiple languages while aligning prediction times.
    \item Exploring the popularity distribution and the correlation between popularity at different times. Based on these, We find the importance of early popularity for popularity prediction task, and point out that the key-point for predicting popularity is to accurately predict the early popularity.
\end{itemize}

\section{Related work}
\textbf{Statistical and topological methods}.
Szabo \textsl{et al.} \cite{szabo2010predicting} found that the early-stage popularity notably influences subsequent popularity. 
Yang \textsl{et al.} \cite{yang2011patterns} conducted cluster analysis on temporal patterns within online content, revealing distinct characteristics of popularity variations across different clusters.
Richier \textsl{et al.} \cite{richier2014bio, richier2014modelling} proposes bio-inspired models to characterize the evolution of video view counts. 
Wu \textsl{et al.} \cite{wu2023predicting} suggested that the information cascade process is best described as an activate–decay dynamic process.

\par
\noindent\textbf{Multi-modal feature based Methods}.
Ding \textsl{et al.} \cite{ding2019social} use pre-trained ResNet and BERT to extract visual and textual features, with the DNN regression. 
Xu \textsl{et al.} \cite{xu2020multimodal} and Lin \textsl{et al.}  \cite{lin2022social} adopted attention mechanisms to effectively integrate multi-modal features.
Chen \textsl{et al.} \cite{chen2019social} compared the performance of several regression models, among which XGBoost \cite{chen2016xgboost} exhibited the best results.
Hsu \textsl{et al.} \cite{hsu2023gradient} employed LightGBM \cite{ke2017lightgbm} and TabNet \cite{arik2021tabnet} to capture intricate semantic relationships in multi-modal features.
Lai \textsl{et al.} \cite{lai2020hyfea} engineered handcrafted features, exploiting CatBoost for regression. Mao \textsl{et al.} \cite{mao2023enhanced} enhanced CatBoost-based model by stacking features.
Tan \textsl{et al.} \cite{tan2022efficient} extracted visual-textual features by ALBEF \cite{li2021align} and Chen \textsl{et al.} \cite{chen2022title, chen2023double} enriched more intermodal features to promote predictive performance.
Wu \textsl{et al.} \cite{wu2022deeply} emphasized increasing feature dimensions to improving predictive performance. The post dependencies captured by sliding window average \cite{wang2020feature} and DSN \cite{zhang2023improving} has also led to improvements. Many of these multi-modal methods are based on the SMPD and working out great \cite{wu2023smp, liu2022review}.

\par
\noindent\textbf{Social media popularity datasets}.
Mazloom \textsl{et al.} \cite{mazloom2016multimodal} conducted experiments using a dataset of posts related to fast-food brands collected from Instagram.
Sanjo \textsl{et al.} \cite{sanjo2017recipe} provided a recipe popularity prediction dataset based on Cookpad, which includes text content entirely in Japanese.
Li \textsl{et al.} \cite{li2019senti2pop} predicted the future popularity of topics by using historical sentiment information based on Twitter's text data records.
TPIC17 \cite{wu2017sequential} and SMPD \cite{wu2019smp} are datasets for single popularity prediction task based on Flickr.
\par
We believe that the loss of temporal alignment about popularity in existing methods is a common problem. The image dynamic popularity dataset \cite{ortis2019prediction} gave us insights, so we propose a multi-modal temporal popularity benchmark to address the shortcomings of existing studies.

\section{SMTPD Dataset}
With the great development of web communication technologies, social multimedia has become the most popular media around the world. 
However, most existing social media prediction datasets are built on the basis of single-output prediction, meaning they observe existing posts and infer their popularity based on their posting time. As mentioned earlier, these data without aligned posting times exhibit non-uniform popularity distributions.
Therefore, SMTPD keep an eye on one of the most popular worldwide social media, YouTube \cite{ceci2022youtube}, with over 282K samples, primarily focusing on the evolution of popularity over time for these samples.
We collected over 402K raw data samples. Specifically, we first removed records with missing values—which likely resulted from network issues during data acquisition. Next, we eliminated samples that either failed to be crawled on certain days or had been deleted within 30 days. Finally, we filtered out potential outliers using the 3$\sigma$ rule.
In this section, we provide a comprehensive overview of the SMTPD's composition. Additionally, we present various data analysis results that contribute to the feasibility of our approach.
\par
Unlike the retrospective methods used in previous datasets, we take note of the newly posted multi-modal content and then observe popularity information every 24 hours via YouTube API.  
Considering that in real-world applications, the posts to be predicted are always new, this data observation method allows us to obtain aligned temporal popularity of new posts, which is more in line with practical applications.
\begin{table}[b]
\centering
\renewcommand\arraystretch{1.5}
\resizebox{\columnwidth}{!}{%
\begin{tabular}{ccc}
\toprule
\textbf{Statistics}  & \textbf{SMTPD} & \textbf{SMPD (Train/Test)}\\
\midrule
Number of samples & 282.4K &  305K/181K\\
Number of users & 152.7K & 38K/31K\\
Mean popularity & 5.95 & 6.41/5.12\\
STD popularity & 4.15 & 2.47/2.41\\
\midrule
Number of categories & 15 & 11 \\
Number of custom tags & 960K & 250K\\
Average length of title & 53.4 chars & 29 words\\
Mean duration & 1853.4 s & - \\
\bottomrule
\end{tabular}%
}
\caption{The basic statistics of all samples in SMTPD, comparing to SMPD\cite{wu2019smp}. Due to the multilingual environment, we use chars (characters) to describe the average length of title.}
\label{table:dataset}
\end{table}
Figure \ref{fig:sample} show a sample from SMTPD. We note diverse attributes, including visual content, textual content, metadata, and user profiles. Additionally, we focus on the temporal popularity for each sample within 30 days after release. We perform basic statistics for SMTPD, as shown in Table \ref{table:dataset}.

\subsection{Temporal Popularity}
The utilization of temporal popularity data is versatile. It not only serves as the output for predictions but also as inputs, where a segment preceding a specific time point, coupled with multi-modal content, can forecast subsequent popularity.
We analyze the distribution and the correlation of popularity over time, so as to facilitate a comprehensive understanding of the significance of temporal popularity for predictions from different perspectives.

A number of popularity definitions have been born from previous work \cite{Li2016analysis, Zo2018}. We use
Khosla's popularity score transformation \cite{khosla2014makes} as it can normalize the distribution of view counts to a suitable popularity score's distribution. It can be represented as:
\begin{equation}
p=\log_{2}({\frac{v}{d}+1})
\label{equ:pop}
\end{equation}
where $p$ is the popularity score, $v$ is the view counts of a sample, $d$ represents the corresponding number of days after the post's release. 
As shown in Figure \ref{fig:distribution}, the histogram of view counts reveals an extreme long-tailed distribution, which might not be suitable as a prediction target. It's evident that the distribution of the popularity score metric becomes more reasonable.

\par
In addition, we notice that there is a strong correlation between the popularity scores of different days. The Pearson Correlation (PC) and Spearman Ranking Correlation (SRC) can respectively reflect the linear relationship and rank-order correlation between popularity scores of different days as follows:


\begin{equation}
PC = \frac{\sum_{i=1}^{n}(X_i - \bar{X})(Y_i - \bar{Y})}{\sqrt{\sum_{i=1}^{n}(X_i - \bar{X})^2}\sqrt{\sum_{i=1}^{n}(Y_i - \bar{Y})^2}}
\label{equ:A}
\end{equation}

\begin{equation}
SRC = \frac{1}{n-1} \sum_{i=1}^{n} (\frac{X_i - \bar{X}}{\sigma_{X}})(\frac{Y_i - \bar{Y}}{\sigma_{Y}})  
\label{equ:AF}
\end{equation}
where $X$ and $Y$ denote the popularity scores belonging to 2 different days, while $\bar{X}$ and $\sigma_{X}$ denote the mean and standard deviation of $X$, and similarly for $Y$. PC and SRC respectively reflect the linear relationship and rank-order correlation between popularity scores of different days. 

\par
The Pearson Correlation (PC) and Spearman Rank Correlation (SRC) heat maps illustrate internal relationships in daily popularity scores, establishing a basis for prediction based on these temporal patterns. Further details are available in the supplementary materials.

\subsection{Multi-Modal Content}
Multi-modal feature fusion is the mainstream approach in today's popularity prediction methods. In this section, we will divide the data into corresponding modality and elucidate the multi-modal content in SMTPD.

\subsubsection{Visual Content}
SMTPD samples, primarily drawn from a multi-modal social media platform, are heavily influenced by visual elements. Since metrics like view count and popularity score are often based on the cover frame that viewers see before clicking to play, we use the cover frame as the primary visual content. This approach balances popularity impact with copyright considerations.

\subsubsection{Textual Content}
The textual content in posts plays a crucial role in media dissemination and user engagement. We examine several core elements, including category, title, description, and hashtags (denoted by the “\#” symbol), along with the user ID (user nickname) from profile metadata. YouTube defines 15 distinct video categories, all in English, which support content organization and discovery. Figure \ref{fig:category} shows the statistical distribution of content by category.

\begin{figure}[H]
    \centering
    \begin{minipage}{0.49\linewidth}
        \centering
        \includegraphics[width=\linewidth]{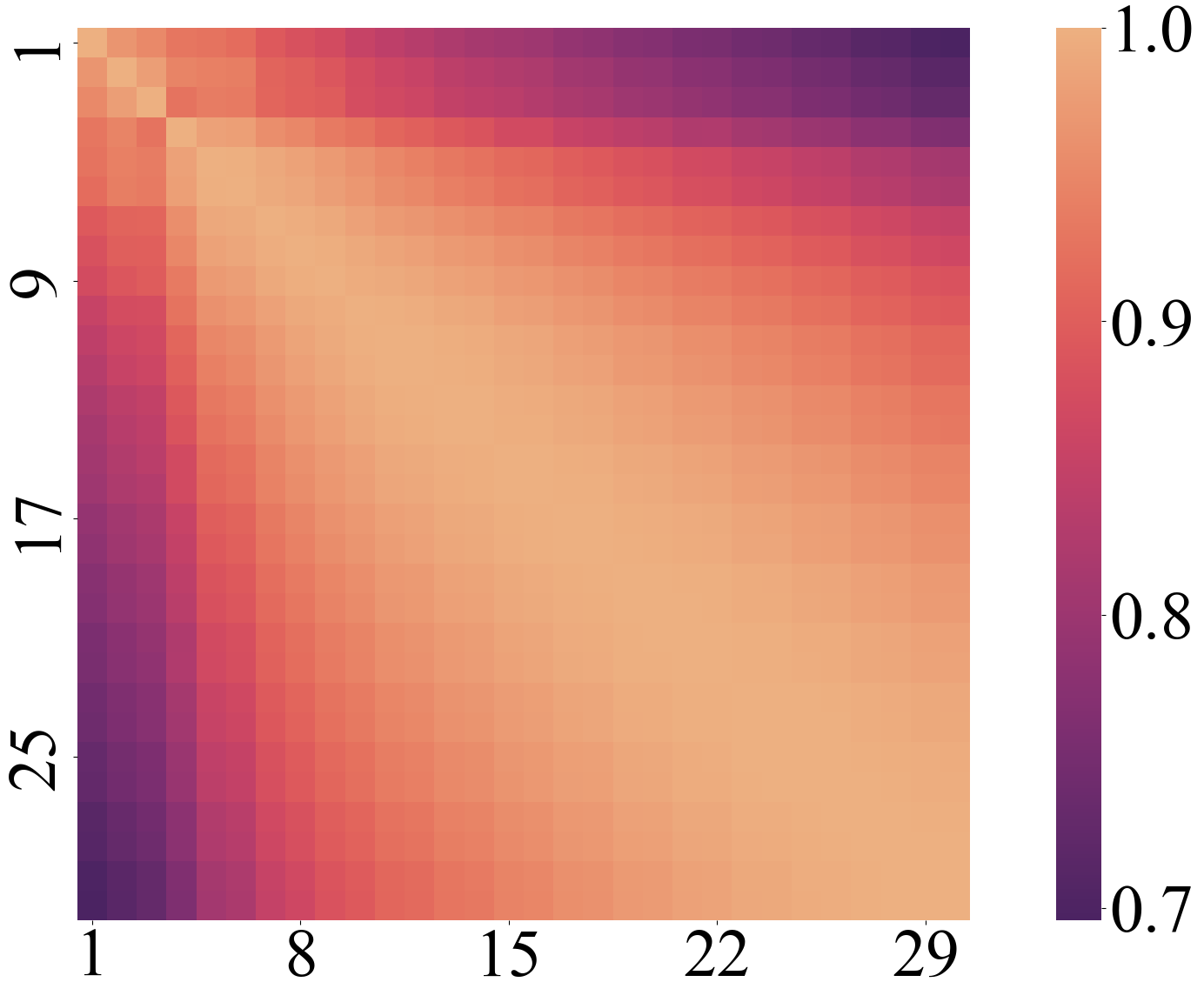}
        \subcaption{The PC Matrix\label{fig:m1}}
    \end{minipage}
    \hfill
    \begin{minipage}{0.49\linewidth}
        \centering
        \includegraphics[width=\linewidth]{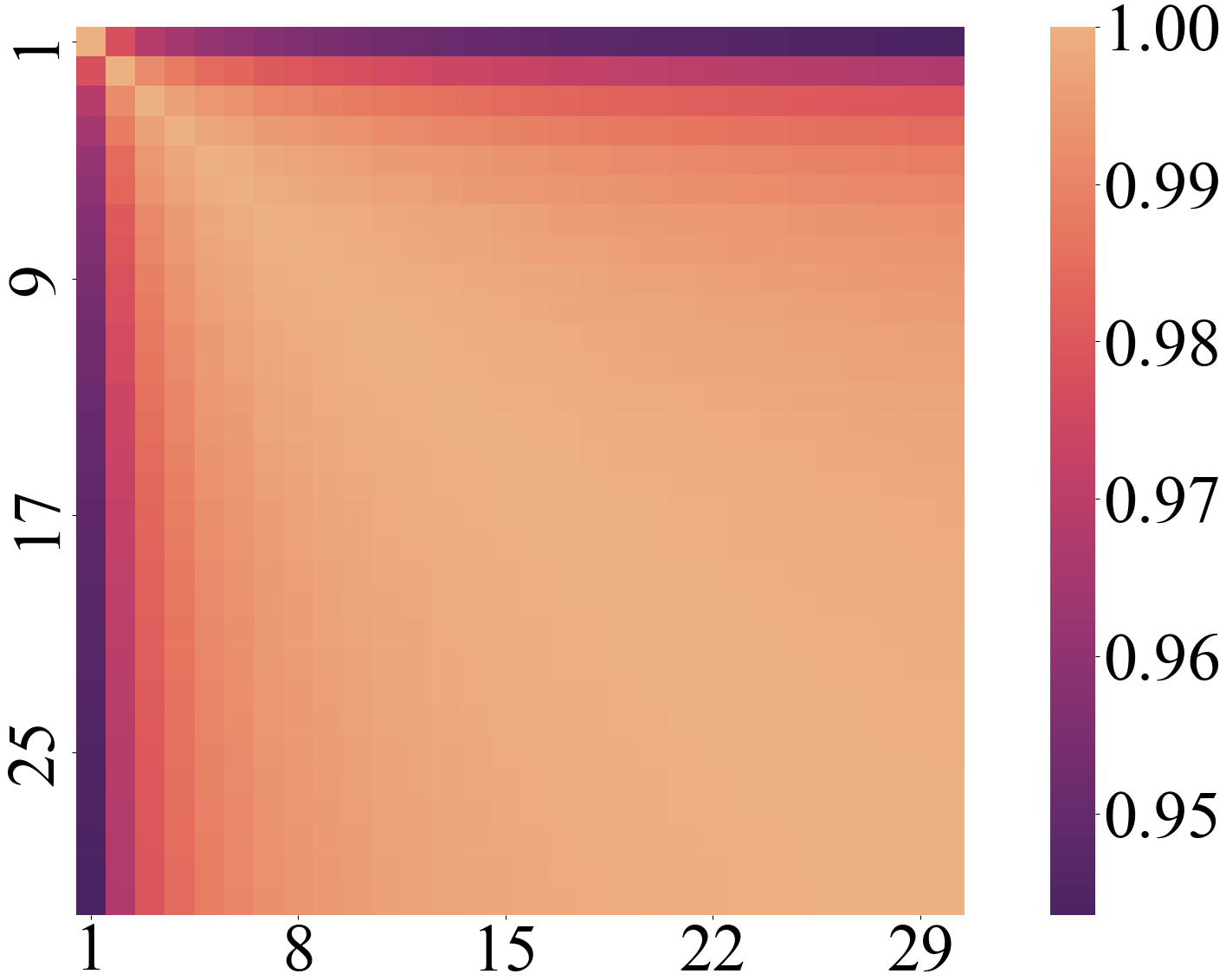}
        \subcaption{The SRC Matrix\label{fig:m2}}
    \end{minipage}
    \caption{\textbf{The heat map of daily popularity correlations.} The figure clearly shows that there is a high degree of correlations in popularity between consecutive days.}
    \label{fig:heat}
\end{figure}

As an international platform, YouTube hosts user-defined textual content in multiple languages beyond standard categories. Figure \ref{fig:languages} illustrates the distribution and popularity bias across title languages.


In previous researches, the datasets typically had uniform language for text content, without considering multilingual aspects. While this simplifies modeling for prediction methods, it also poses significant limitations for international social media platforms which not restricted to a single language. Besides, predictions based on a few languages may introduce biases \cite{ghosh2021detecting} against popularity.

\subsubsection{Numerical values}
Numerical attributes frequently function as supplementary factors in enhancing prediction accuracy. In this study, we examined key attributes, including video duration, uploader follower count, and the number of posts by the uploader. Additionally, we manually derived several auxiliary metrics, such as title length, the count of custom tags, and description length, to further enrich the feature set and support predictive robustness.

\section{Temporal popularity prediction}
The multi-modal feature-based temporal prediction framework is shown in Figure \ref{fig:model}, which is divided into two main parts: multi-modal feature extraction and temporal popularity score regression.
This framework is designed to adapt to the multilingual temporal prediction tasks under time alignment. 
The prediction target is set as the popularity of samples, which corresponds to the temporal popularity within 30 days after the sample's release.

\begin{figure}[b]
    \centering
    \subfloat[The statistics of samples in each category]{\includegraphics[width=1\linewidth]{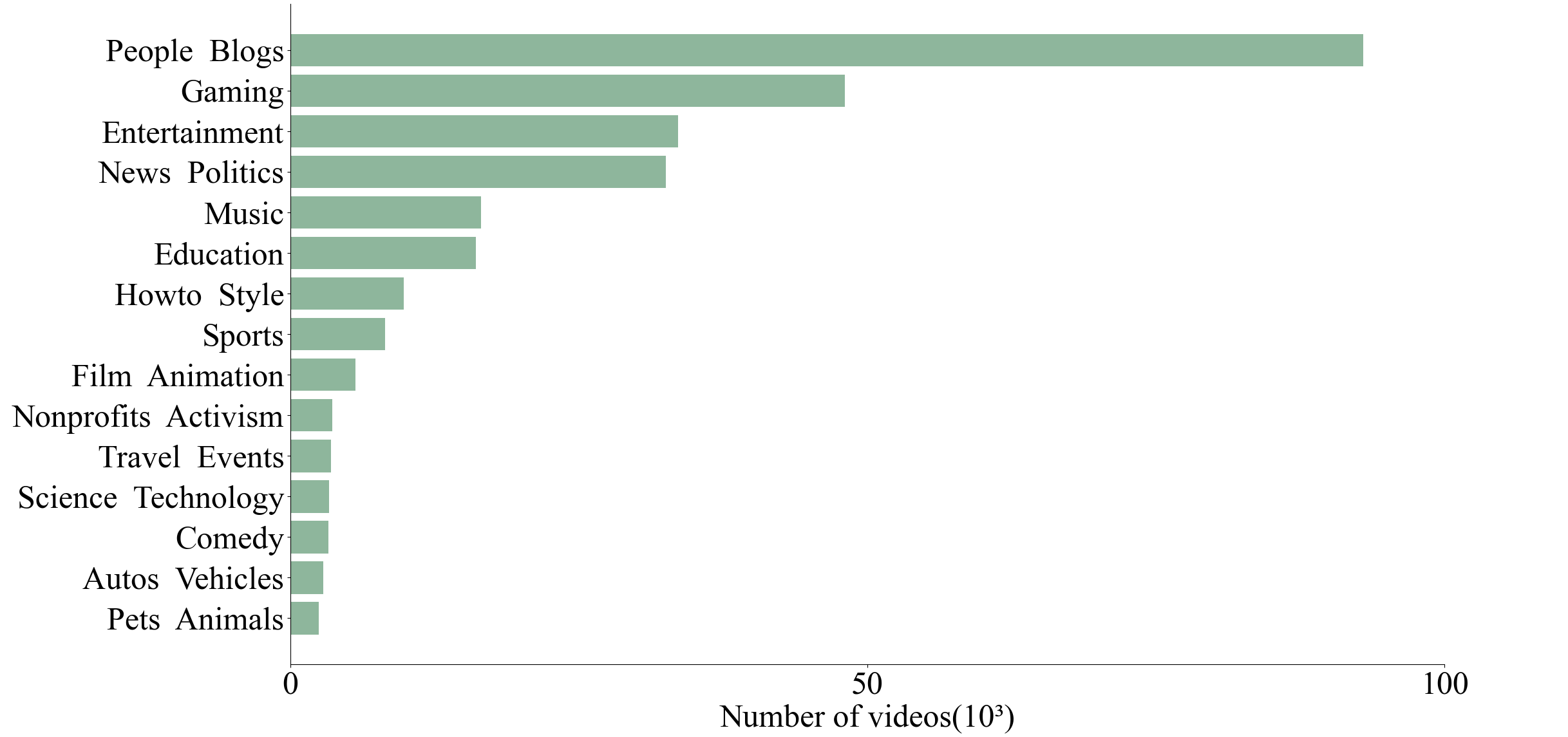}\label{fig:c1}}\\
    \subfloat[Average popularity of each category]{\includegraphics[width=1\linewidth]{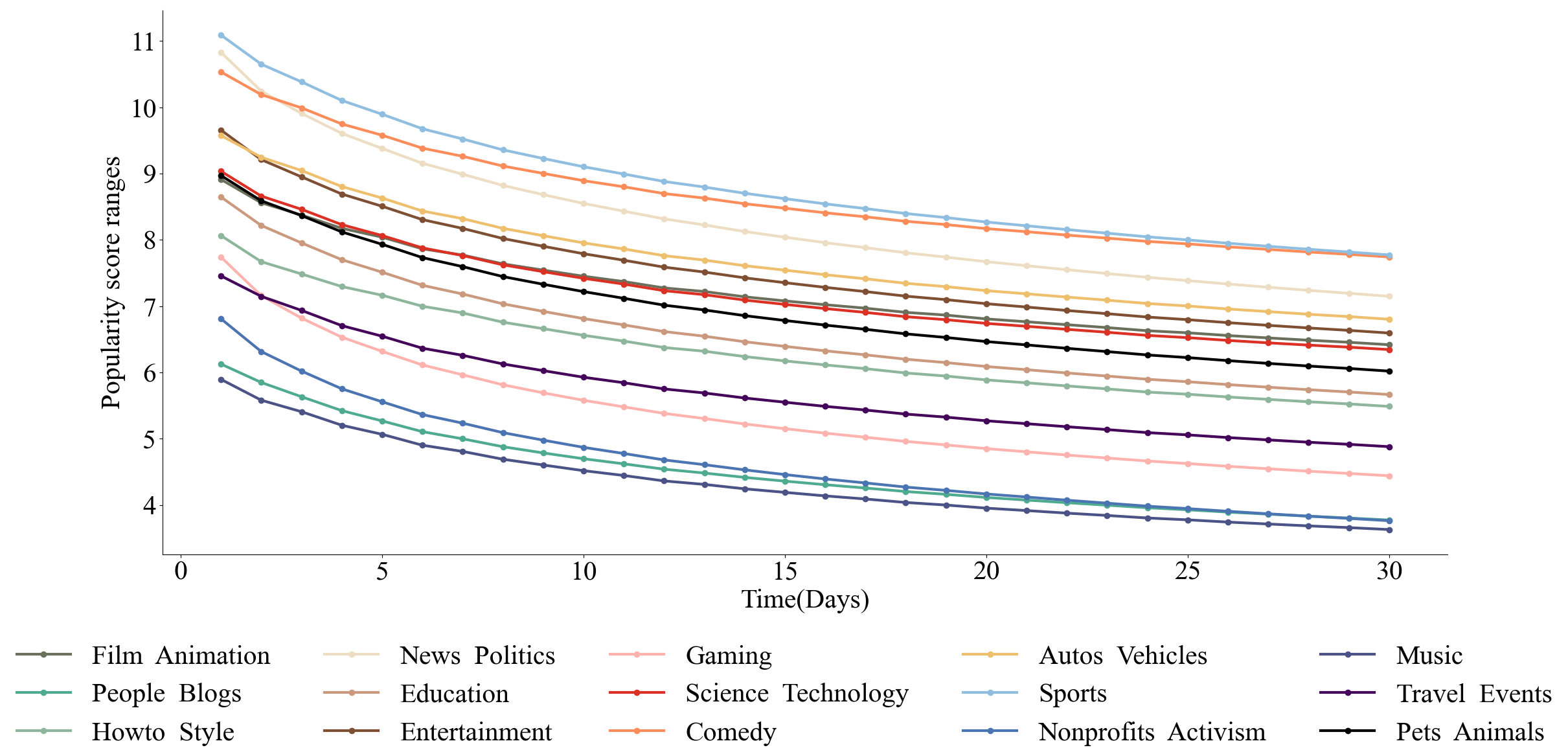}\label{fig:c2}}
    \caption{\textbf{The statistics based on category.} \ref{fig:c1} counts the number of samples in each category, and \ref{fig:c2} shows the average popularity score of samples in each category.}
    \label{fig:category}
\end{figure}


\begin{figure}[t]
    \raggedright
    \centering
    \includegraphics[width=1\linewidth]{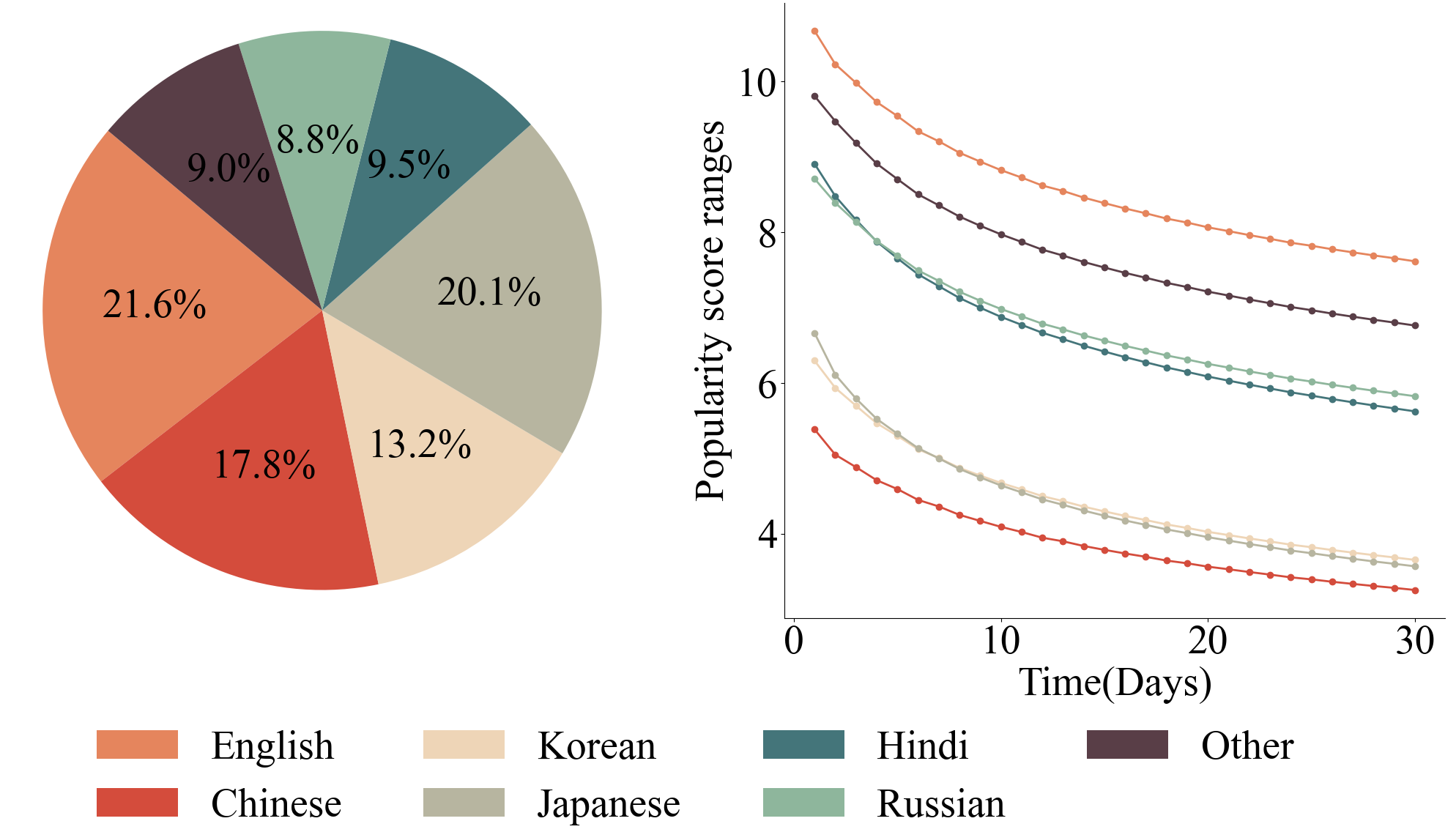}
    \caption{\textbf{Languages analysis.} The left is the proportions of these languages, with "Other" encompassing 90 different languages. The right represents the average popularity in different languages, revealing the geographic biases . The languages of samples are counted by reference to the title.}
    \label{fig:languages}
\end{figure}

\subsection{Multi-Modal Feature Extraction}
For different modalities in SMTPD, we adopted distinct feature extraction methods. 
We additionally incorporated features from the categorical modality to investigate the impact of categorical features on popularity prediction.

\subsubsection{Visual Features}
The cover image of a social media content is a very important component which would help users to quickly understand what they would see if clicking to view this content.
Therefore, the cover image would directly affect the popularity of a social media content.
The semantic information provided in the cover image plays a key role for user's understanding.
We adopt convolutional neural networks ResNet-101 \cite{he2016deep} pre-trained on ImageNet as our visual feature extraction model to obtain the semantic information provided by the cover.
To match the input size of ResNet-101, we re-scale the size of cover image to $224\times224$.
The 2048-dimension feature vector before the final classification layer will be selected as the final visual feature $f_v$:
\begin{equation}
f_v=\text{ResNet}(\mathcal{S}(I))
\label{equ:visual}
\end{equation}
where $I$ denotes the cover image and $\mathcal{S}$ denotes the re-scale operation.
However, the cover image would not always be uploaded by authors or be unavailable due to the network transmission. 
For those contents missing the cover image, we use blank images with all pixels are set to zero instead. 

\subsubsection{Textual Features}
Textual information significantly impacts a user’s choice to view or skip content. Key textual inputs include category, title, tags, description, and user profile ID. Extracting semantic features from these inputs is crucial for accurate popularity prediction.

However, unlike previous popularity prediction tasks, one distinguishing characteristic of SMTPD is that its text content includes multiple languages, and many samples contain text in more than one language. 
This makes it challenging to use many pre-trained word vector models based on single-language corpora, such as \cite{grave2018learning, mikolov2013efficient, pennington2014glove}.

Thanks to the multilingual capability of BERT-Multilingual \cite{devlin2018bert}, which processes text across languages in a single model, we use it to extract a 768-dimensional feature vector for each text, capturing semantic information as:
\begin{equation}
    f^{k}_{t}=\text{BERT}(T^{k})
\end{equation}

\begin{figure}[t]
    \centering
    \setlength{\belowcaptionskip}{1pt}
    \includegraphics[width= 0.9\linewidth]{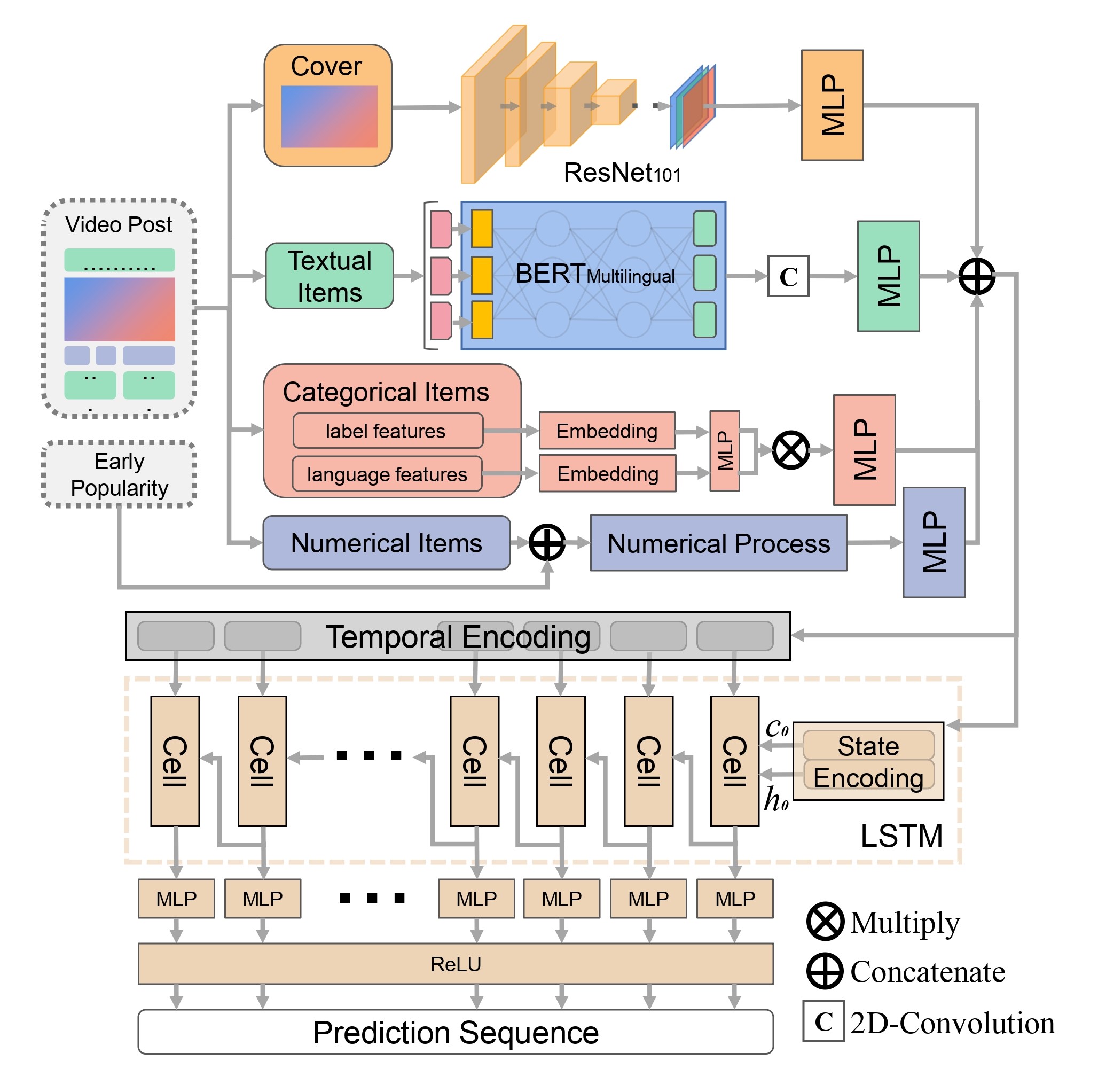} 
    \caption{\textbf{The proposed model consists of two layers.} The upper layer processes visual, textual, numerical, and categorical features, which are input to a multi-layer perceptron (MLP). The MLP outputs are concatenated and fed into a lower LSTM layer for regression. Each LSTM cell's output is passed through another MLP, with the final MLP outputs combined to generate a time prediction sequence using the ReLU activation function.}
    \label{fig:model}
\end{figure}

where $k\in\{cat, tit, tag, des, uid\}$ maps to category, title, tags, description, and user ID. These are combined into a $768\times5$ matrix, transformed through $5\times5$ convolution to produce the 768D textual feature $f_t$, which can be expressed as:
\begin{equation}
    f_{t} = \text{Conv}_{5\times5}(\mathcal{C}(f^{cat}_{t},f^{tit}_{t},f^{tag}_{t},f^{des}_{t},f^{uid}_{t}))
\end{equation}
where $\mathcal{C}$ denotes the above combing operation.

\subsubsection{Numerical Features}
Numerical metrics directly influence whether an audience clicks to watch which sample on social media platform. 
Also, the number of followers and the number of posts also impact the process through which an audience navigates from content uploaders' user pages to click and watch videos. Hence, we selected these influential numerical metrics for numerical features extraction.As the number of user followers share a similar distribution pattern that exhibiting long-tailed characteristics as view counts, it is necessary to apply a logarithmic transformation before feeding it into the network. Afterward, all numerical items are Z-score normalized and concatenated as numerical features:
\begin{alignat}{2}
    &f^j_n = \frac{T^j - \mu_{j}}{\sigma_{j}}  \\
    f_n = \text{Concat}
            (f^{fol}_{n}&,f^{pos}_{n},f^{dur}_{n},
            f^{tl}_{n},f^{tn}_{n}, f^{dl}_{n}, f^{EP}_{n})            
\end{alignat}
where $j \in \{fol, pos, dur, tl, tn, dl, EP\}$ represents the log-scaled follower counts, total number of posts, video duration, title length, the number of tags, description length and the early popularity (the 1st day’s popularity). The $\mu$ and $\sigma$ represents the mean and standard deviation of all samples. And Concat denotes the horizontal concatenating operation.

\subsubsection{Categorical Features}
The categorical disparities across social media platforms are depicted in Figure \ref{fig:c2}. Beyond text-based feature extraction, these disparities inform the generation of categorical features. Language functions as a key marker of regional and cultural distinctions among content creators, as highlighted by the language bias observed in Figure \ref{fig:languages}. Consequently, language classification features are crucial for enhancing prediction accuracy. To develop these classification features, two sub-features are selected: label and language attributes. The label feature captures the semantic category, while the language attribute is derived from the language in the video title. Assuming that the primary language of all text content matches the title’s language, we use langid \cite{lui2012langid} for classification. Two embedding layers are introduced to encode label and language features independently, each processed by a dedicated multi-layer perceptron (MLP). Finally, the outputs are combined via cumulative multiplication to form the classification feature $f_c$, defined as:
\begin{equation}     
f_{c} = \text{MLP}_{1}(\text{E}_{1}(T^{cat})) \odot \text{MLP}_{2}(\text{E}_{2}(\text{langid}(T^{tit}))) \label{equ:cat} 
\end{equation}

where $\text{E}_1$ and $\text{E}_2$ represent two independent embedding layers, and $\text{MLP}_1$ and $\text{MLP}_2$ refer to multi-layer perceptrons applied to the label embedding $\text{E}_1(T^{\mathit{cat}})$ and the language embedding $\text{E}_2(\text{langid}(T^{\mathit{tit}}))$, respectively. The operator $\odot$ denotes element-wise multiplication, which combines the outputs of the two MLPs to produce the final classification feature $f_c$.

\subsubsection{Feature Fusion}
After extraction, features from different modalities are aligned in dimension through multi-layer perceptrons (MLPs) and then concatenated into fused features $F$.

\subsection{Temporal Popularity Regression}
\subsubsection{Sequential Encoding and Adjustment}
Considering the high correlation between popularity values on adjacent days in the ground truth, we employ a LSTM \cite{hochreiter1997long} structure to get the temporal variations in popularity.
As shown in Figure \ref{fig:model}, multiple LSTM cells are constructed to transmit temporal information. 
The initial states (both hidden state and cell state) of the LSTM are generated by passing the $F$ through the same MLP as state encoding. 
Additionally, the input for LSTM cell at each time step is constructed from $F$ through temporal encoding MLPs. These two encoding process can be written as:
\begin{alignat}{2}
h_0 = c_0& = \text{MLP}^{\text{hc}}(F)\\
x_s = &\ \text{MLP}^{\text{x}}_s(F) 
\end{alignat}
In this formulation, $h_0$ and $c_0$ denote the initial hidden state and cell state, while $x_s$ represents the input at time step $s$. The $\text{MLP}^{\text{hc}}$ module encodes the state information, and $\text{MLP}^{\text{x}}_s$ handles the temporal encoding at the $s$-th time step, effectively capturing the sequence dynamics.
\par
At each time step, both states are treated as the $s$-th step's features for output. After concatenating and processing through independent MLPs, each LSTM cell produces outputs, helping capture temporal popularity via backpropagation during training. Before final predictions, we apply a non-negative adjustment since popularity scores cannot be negative. This process is formulated as:
\begin{equation}
pre_s = \max(0, \text{MLP}^{\text{out}}_s(\text{Concat}(h_s, c_s)))
\label{equ:LSTM}
\end{equation}
Where $pre_s$, $h_s$, $c_s$, and $\text{MLP}^{\text{out}}_s$ denote the predicted value, hidden state, cell state, and output MLP at time step $s$, respectively. The operation $\text{Concat}$ refers to the process of horizontally concatenating the vectors. After these steps, the final temporal popularity predictions are produced.

\subsubsection{Loss Function}
The core component of our approach is the Composite Gradient Loss (CGL), specifically designed for this task. This custom loss function consists of several components: SmoothL1Loss (SL) between the model's outputs and targets, the first-order and second-order derivative differences between the outputs and targets, the L1 loss between the onehot encodings of the predicted and ground truth peaks, and the Laplacian remainder (LR). These components are combined with a weight ratio of 1:1:1:1e-6. The overall loss function $\mathcal L$ is formulated as:
\begin{small}
\begin{align}
    \mathcal L = & \ \text{SL}(\hat{P}_{d,i}, P_{d,i}) 
     + \lambda_{1} \cdot \text{SL}(\hat{P}_{d,i}^{(1)},P_{d,i}^{(1)})  
     + \lambda_{2} \cdot \text{SL}(\hat{P}_{d,i}^{(2)},P_{d,i}^{(2)})  \\
    & + \alpha \cdot \sum_{i=1}^{n} \left| \delta \text{argmax}_{d} (\hat{P}_{d,i}) - \delta \text{argmax}_{d} (P_{d,i}) \right| \nonumber + \epsilon \cdot \mathrm{LR}
\end{align}
And the Laplacian remainder is:
\begin{align}
    \mathrm{LR} = \sum_{i=1}^{n} \left| \hat{P}_{d,i}^{(1)} \right| + \sum_{i=1}^{n} \left| \hat{P}_{d,i}^{(2)} \right|
\end{align}
\end{small}
The term $\text{SL}(\hat{P}_{d,i}, P_{d,i})$ (with $\beta = 0.1$) is used to compute the error between the predicted outputs and the ground truth targets. Here, $\hat{P}_{d,i}$ represents the predicted popularity for data point $i$ on day $d$, while $P_{d,i}$ denotes the ground truth popularity for the same data point. The differences between the one-hot encoded predicted and true peak values are measured using $\delta \text{argmax}_{d} (\hat{P}_{d,i})$ and $\delta \text{argmax}_{d} (P_{d,i})$, respectively. Additionally, the term $\epsilon \cdot\mathrm{LR}$ introduces a Laplacian remainder (LR) to provide further regularization.
During training, the weights for the first-order and second-order derivative terms, as well as the L1 loss weight between the one-hot encodings ($\lambda_{1}$, $\lambda_{2}$, $\alpha$), are adjusted dynamically using a cosine annealing algorithm to ensure smoother convergence throughout the training process.
\section{Experiments}
In this section, we first describe the experiment settings and evaluation metrics for training models and comparison. 
Then various experimental results and detailed discussions are provided including evaluations for SMTPD dataset, proposed baseline, multi-modal features and early popularity. 

\subsection{Experiment settings}
We implement our approaches using the Pytorch\footnote{\hyperlink{pytorch-site}{https://pytorch.org}} framework and train it with the Adam \cite{kingma2014adam} optimizer incorporating L2 penalty of $10^{-3}$, while the batch size is set to 64. 
The learning rate is initialized to $10^{-3}$ and adjusted by the ReduceLROnPlateau scheduler provided by PyTorch when one epoch is end. Supervised training was performed using the Composite Gradient Loss (CGL) metioned before. 
\par
We introduce error-based metrics and correlation-based metrics to fairly evaluate the compared datasets and models. We introduce the Absolute Error (MAE) and average MAE (AMAE) to respectively evaluate models for single-day prediction and temporal prediction.
Assuming that the daily MAE and average MAE are denoted as $MAE_d$ and $AMAE$, they can be defined as:
\begin{alignat}{2}
    MAE_d = \frac{1}{n} \sum_{i=1}^{n} &\left|\hat{P}_{d,i} - P_{d,i} \right|  \\
    AMAE = \frac{1}{m} &\sum_{d=1}^{m} MAE_d  
\end{alignat}
where $n$ denotes the total of samples and $m$ denotes total of days, while the $P_{d}$ and $\hat{P}_{d}$ respectively denotes the ground-truth popularity and predicted popularity for the day $d$.
\par
For correlation-based metrics, we introduce the daily SRC and the average SRC to evaluate the models for both single-day prediction and temporal prediction from a different perspective. Assuming $n$, $m$, $P_{d}$, and $\hat{P}_{d}$ are defined consistently as mentioned earlier, the daily SRC and average SRC can be formulated as follows:
\begin{alignat}{2}
    SRC_d = \frac{1}{n-1} \sum_{i=1}^{n} &(\frac{\hat{P}_{d,i} - \bar{\hat{P}}_d}{\sigma_{\hat{P}_d}})
(\frac{P_{d,i} - \bar{P}_d}{\sigma_{P_d}})   \\
    ASRC = &\frac{1}{m} \sum_{d=1}^{m} SRC_d  
\end{alignat}
where $\bar{P_d}$ and $\sigma_{P_d}$ are mean and standard deviation of the corresponding popularity for the day $d$.

In the curves, sequential performance is represented by the MAE and SRC metrics, which illustrate daily prediction errors and correlations for popularity scores. The X-axis denotes days, while the Y-axis shows $MAE_d$ 
or $SRC_d$.

\subsection{SMTPD VS. SMPD}

In this subsection, we conduct comparisons and discussions between the SMTPD and SMPD.
We first measure three most recently proposed state-of-the-art popularity prediction methods on two datasets. 
As these methods are all designed basing on the settings of SMPD, we make some modifications to let them fit for training and testing on our SMTPD. Other top-performing models (e.g. \cite{wang2020feature, tan2022efficient, wu2022deeply}) include components that are not applicable to SMTPD or are not reproducible, such as sliding window average or undisclosed self-trained modules. Hence, we do not discuss them in our experiments.
First, we distribute content from corresponding modalities into their respective feature extraction modules. 
Then we use BERT-Multilingual as the textual feature extractor to address the challenges of multilingual text. Given that our SMTPD dataset consists of temporal popularity scores, we restrict these methods to predict only the single popularity score on day 30. To mitigate random bias, we evaluate using $5$-fold cross-validation.


Table \ref{table:ref} shows minimal performance differences across folds, confirming that SMTPD contains abundant data under a well-balanced sample distribution. 
It is evident that on the single-output prediction of popularity on days 7, 14, and 30 within SMTPD, the performance of method \cite{lai2020hyfea} based on GBDT outperforms the deep learning method \cite{ding2019social} and \cite{xu2020multimodal} in terms of both MAE and SRC through its powerful regression capabilities of Catboost \cite{prokhorenkova2018catboost}. Notably, our method achieves the best results with the addition of EP, where the AMAE of the other three models increases as the prediction horizon extends. In contrast, our model’s AMAE begins to decrease after day 14. The deep learning method \cite{ding2019social} also performs well on days 7 and 14, but its effectiveness declines beyond day 14.

Morever there are also two interesting observation. First, all methods achieve higher SRC on SMTPD than on SMPD, likely because SMPD lacks specific time points, making it challenging to predict time-dependent popularity trends accurately as popularity declines and their MAE performances are certainly reduced that the MAE values increase by over 0.16. Besides, the larger popularity range and standard deviation among SMTPD samples (shown in Table \ref{table:dataset}) also contributes to prediction difficulties.

\begin{table}[b]
\centering
\renewcommand\arraystretch{1.5}
\resizebox{\columnwidth}{!}{%
\begin{tabular}{cccccc}
\toprule
BERT-Base & BERT-Mul & MLP & LSTM & AMAE & ASRC \\
\midrule
\checkmark & &  &\checkmark & 0.782 & 0.958 \\
 & \checkmark & \checkmark & & 0.786 & 0.958 \\
 & \checkmark &  & \checkmark & \textbf{0.717} & \textbf{0.959}  \\
 \bottomrule
\end{tabular}%
}
\caption{The evaluation for the proposed baseline model, mainly on evaluating the model's performance in adapting to multilingual and temporal popularity.}
\label{table:abl1}
\end{table}

\begin{table}[h]
\centering
\renewcommand\arraystretch{1.2}
\resizebox{\columnwidth}{!}{%
\begin{tabular}{cccc} 
\toprule
\multirow{2}{*}{\textbf{Method}} & \multicolumn{1}{c}{\textbf{SMTPD (day 7)}} & \multicolumn{1}{c}{\textbf{SMTPD (day 14)}} & \multicolumn{1}{c}{\textbf{SMTPD (day 30)}} \\ 
\cmidrule(lr){2-2} \cmidrule(lr){3-3} \cmidrule(lr){4-4}
& \textbf{Average} & \textbf{Average} & \textbf{Average} \\
\midrule

Ding \textit{et al.} \cite{ding2019social} 
 & 1.715/0.849 & 1.669/0.846 & 1.592/0.843\\

w. EP  & 0.715/0.964 & 0.742/0.959 & 0.749/0.931\\

\midrule

Lai \textit{et al.} \cite{lai2020hyfea} 
 & 1.573/0.875 & 1.524/0.872 & 1.495/0.864\\

w. EP  & 0.725/0.957 & 0.753/0.962 & 0.760/0.957\\

\midrule

Xu \textit{et al.} \cite{xu2020multimodal} 
 & 1.895/0.817 & 1.832/0.818 & 1.743/0.820\\

w. EP  & 0.754/0.962 & 0.798/0.956 & 0.822/0.949\\

\midrule

\textbf{Ours w/o. EP} 
 & 1.673/0.852 & 1.628/0.850 & 1.563/0.848\\

\textbf{Ours} 
 & \textbf{0.713/0.964} & \textbf{0.735/0.959} & \textbf{0.732/0.959}\\

\bottomrule
\end{tabular}%
}
\caption{The performance (MAE/SRC) was compared across four models, including our model, using the SMTPD dataset, both with and without EP.}
\label{table:ref}
\end{table}

\subsection{Evaluation for Proposed Baseline}
Refer to Table \ref{table:abl1}, we attempted to validate the rationale of partial structures in the proposed baseline model  by employing alternative feature extractors and regression networks. Using BERT-Base as the textual features extractor lead to an increase of MAE by around 0.19. This decline in performance may be caused by the limitations of BERT-base model in handling multilingual texts, as it struggles to effectively capture the semantics and contextual information across different languages. The MAE also increased when the MLP (3 layers) was used as the regression structure. Such a structure is similar to \cite{ding2019social}, suggesting that MLP lacks the capture of temporal information. The SRC did not change for either of the above substitutions, probably due to the high correlation given by EP.

\par

\subsection{Discussion of Early Popularity}
We evaluate the performance of proposed baseline on SMTPD.As shown in Table 5, the proposed baseline gets a great improvement from existed methods that surpassing the second best method by -0.798/0.109
MAE/SRC for predicting the popularity of day 30.
This great improvement is mainly brought by the early popularity (EP). Without the assistant of EP (row 2 in Table \ref{table:result1}), the baseline performance sharply reduced around the existed methods.Using the EP predicted by existed method (1.717/0.864 MAE/SRC) as an input also contributes to the baseline's performance (row 3 in Table \ref{table:result1}).However the performance gap between the predicted EP and true EP remains large.

Instead of directly involving the EP in the model architecture, introducing EP in the training supervision is another suitable option.To validate the effectiveness of such operation, we train a baseline model for predicting the popularities from day 1 to 30. As it presented by the row 1 in Table \ref{table:result1}, adding the 1st day's popularity to the prediction sequence make a slight boost, what is in line with the LSTM's ability to capture temporal dependencies in popularity sequences. Having more preceding temporal information from backpropagation leads to more precise predictions of popularity in subsequent time steps.

These comparisons clearly indicate that EP plays a crucial role in popularity prediction, and accurately forecasting the first day’s popularity is key to predicting future popularity. The strong correlation between EP and future popularity significantly benefits the prediction task. Our results, as demonstrated by the MAE and SRC curves for both natural EP and our model’s predictions, underscore the model’s capability to leverage EP for enhanced accuracy. From the visualized results, tough EP exhibits high correlation to the popularities of following days, the MAE sharply increased over time. By contrast, our baseline model could well optimize the MAE and achieve even better SRC performance comparing to the natural EP curve.
Therefore, the proposed baseline model is effective to utilize EP achieving better prediction accuracy.

\begin{table}[t]
\centering
\renewcommand\arraystretch{1.5}
\resizebox{\columnwidth}{!}{%
\begin{tabular}{ccccc}
\toprule
\multirow{2}{*}{\textbf{Method}} & \multirow{2}{*}{\textbf{AMAE}} & \multirow{2}{*}{\textbf{ASRC}} & \textbf{MAE} & \textbf{SRC} \\
& & & \footnotesize{(day 30 only)} & \footnotesize{(day 30 only)} \\
\midrule
w/o. EP(1-30) & 1.562 & 0.856 & 1.530 & 0.850 \\
w/o. EP(2-30) & 1.630 & 0.849 & 1.551 & 0.849 \\
w/o. EP+\cite{lai2020hyfea} & 1.628 & 0.851 & 1.555 & 0.848 \\
ours & 0.717 & 0.959 & 0.732 & 0.959 \\
\bottomrule
\end{tabular}%
}
\caption{\textbf{Assessment of EP across different scenarios.} Here, "1-30" denotes the prediction target spanning a continuous sequence from the 1st day to the 30th day, as does "2-30" (to align with methods having EP).}
\label{table:result1}
\end{table}
\section{Conclusion}
This study aims to address the challenge of time-series popularity prediction in social media. In this paper, we introduce a novel multilingual, multi-modal time-series popularity dataset based on YouTube and suggest a multilingual temporal prediction model tailored to this dataset. Through experiments, we demonstrate the effectiveness of this approach in predicting social media popularity time-series in a multilingual environment. 
The experiments show that combining multi-modal features with early popularity significantly improves prediction accuracy. However, this study has yet to address challenges such as multi-frame information in videos, maximizing the use of language diversity, and deeper multi-modal exploration. These areas will be key focuses for our future work.


\clearpage
{
 \small 
 \bibliographystyle{ieeenat_fullname} 
 \bibliography{main}
}


\end{document}